# Decoding Thermal Stability: *In situ* Insights into Phase Controlled Phosphine-free Colloidal Bi-Te Nanosheets

*Fa-Gui He,[1*] Kevin Oldenburg,[1,3] Rostyslav Lesyuk,[1,2] and Christian Klinke[1,3*]*

[1] Institute of Physics, University of Rostock, Albert-Einstein-Straße 23, 18059 Rostock, Germany

[2] Pidstryhach Institute for applied problems of mechanics and mathematics of NAS of Ukraine, Naukowa 3b, 79060 Lviv, Ukraine

[3] Center for Interdisciplinary Electron Microscopy (ELMI-MV), Department "Life, Light & Matter", University of Rostock, Albert-Einstein-Straße 25, 18059 Rostock

ABSTRACT

Bismuth telluride ($Bi_2Te_3$) is a prototypical V-VI semiconductor of interest for both thermoelectric and topological applications; however, phase-selective synthesis and thermal stability across the bismuth-telluride homologous series remain poorly understood. We report a facile and phosphine-free colloidal synthesis approach for a phase-selective synthesis of $Bi_2Te_3$, BiTe, and $Bi_4Te_5$ nanosheets with well-defined hexagonal morphology. By controlling precursor chemistry and reaction temperature, we achieved the selective formation of different phases within the $(Bi_2)_m(Bi_2Te_3)_n$ homologous series. Based on *in situ* heating studies, BiTe and $Bi_4Te_5$ nanosheets transform into $Bi_2Te_3$ at ~340 °C, followed by preferential Te sublimation under vacuum or oxidation in air at higher temperatures. We discuss plausible mechanisms for these phase transformations. EDS analysis and FFT analysis of STEM images provide direct evidence for the temperature-dependent compositional and structural changes and highlight the close thermal relationship among these phases. These results not only advance fundamental understanding of phase stability and thermal behavior in bismuth-tellurides at the nanoscale but also establish a framework for understanding structural evolution in related homologous series, providing insights into their potential future applications in thermoelectric, spintronic, and topological systems.

Bismuth telluride ($Bi_2Te_3$), a prominent V-VI semiconductor with a bandgap of 0.15 - 0.2 eV,[1] is widely recognized as one of the most efficient thermoelectric materials near room temperature.[2, 3] Simultaneously, it serves as a three-dimensional topological insulator with potential applications in spintronics and quantum computing.[4] Nanostructuring has been shown to significantly enhance its thermoelectric performance by reducing thermal conductivity through enhanced phonon scattering at grain boundaries and increasing the density of states in confined structures, thereby improving electrical conductivity.[5] Furthermore, electrical transport can be finely tuned via controlled defect activation, while the properties of topologically protected surface states strongly depend on surface termination, highlighting the importance of morphology control for tailoring electronic properties.[5, 6] Consequently, the synthesis of $Bi_2Te_3$ nanostructures with controllable morphology and size has attracted extensive attention. A variety of approaches, including wet-chemical methods, have enabled the fabrication of $Bi_2Te_3$ nanostructures with diverse morphologies, such as nanorods, nanowires, nanosheets, hexagonal nanoplates, rod-like assemblies, and flower-like architectures.[5, 7-14]

The crystal structure of $Bi_2Te_3$ is a rhombohedral tetradymite-type structure with the space group R-3m,[15] typically described by a hexagonal unit cell (Figure 1a). Its atomic arrangement consists of charge-neutral quintuple layers stacked along the *c*-axis in the sequence -Te(1)-Bi-Te(2)-Bi-Te(1)-. Within each quintuple, atoms are primarily covalently bonded with a minor ionic contribution, while adjacent quintuples are weakly connected through van der Waals interactions, resulting in a natural cleavage plane between neighboring Te(1) layers. The superscripts (1) and (2) indicate chemically distinct tellurium sites. This layered configuration leads to pronounced anisotropy: the lattice constant along the *c*-axis (≈ 3.045 nm) is about seven times larger than that along the *a, b*-axes (≈ 0.438 nm).[15, 16] Importantly, the bandgap in a single quintuple is larger than in few-quintuple layers due to the reduced coupling of top and bottom surface states in thinner structures.[1] These intrinsic crystallographic features not only determine the physical properties of $Bi_2Te_3$ but also explain its excellent exfoliation ability and the facile formation of sheet-like nanostructures, which are crucial for both thermoelectric and topological applications.

Despite significant progress in the colloidal synthesis of $Bi_2Te_3$ nanostructures, trioctylphosphine-tellurium (TOP-Te) remains one of the most commonly used tellurium precursors. In our previous work, the TOP-Te route successfully produced SnTe nanosheets, and TOP was found to play a beneficial role in stabilizing the two-dimensional growth template.[17] However, in the present study, we observed notable inconsistencies with our earlier results when using the conventional TOP-Te precursor. Specifically, the synthesized $Bi_2Te_3$ nanosheets were significantly smaller than expected. Moreover, when a small amount of TOP was introduced prior to the injection of TOP-Te in an attempt to better stabilize the two-dimensional growth template, the product was predominantly metallic Bi nanosheets instead of $Bi_2Te_3$. These findings indicate that TOP not only serves as a surface capping ligand but can also strongly promote the reduction of $Bi^{3+}$ under the current reaction conditions. To address this discrepancy, we developed a phosphine-free colloidal approach using commercially available diphenyl ditelluride dissolved in oleylamine as the tellurium precursor. This method enables the synthesis of much larger $Bi_2Te_3$ nanosheets (lateral sizes up to ~10 μm) with improved shape uniformity. Moreover, by simply tuning the reaction temperature, selective synthesis of other Bi-Te phases, namely BiTe (at 230 °C) and $Bi_4Te_5$ (at 240 °C), was also achieved.

Beyond $Bi_2Te_3$, the bismuth-telluride system encompasses a rich variety of binary compounds with different Bi:Te ratios, forming an adaptive homologous series $(Bi_2)_m(Bi_2Te_3)_n$ (where *m* and *n* are integers) within certain compositional limits.[18-20] These phases display significant structural and functional diversity, with at least seven reported crystal structures. Most of them are layered and adopt rhombohedral or trigonal symmetry, which are most frequently represented using hexagonal lattices with nearly identical in-plane lattice constants (~0.44 nm), while their *c*-axis dimensions vary significantly from ~1 to 15 nm.[21, 22] The series includes both widely studied "end members" such as Bi (m:n = 3:0) and $Bi_2Te_3$ (m:n = 0:3), as well as intermediate compounds like BiTe (m:n = 1:2) , $Bi_2Te$ (m:n = 2:1) , $Bi_4Te_5$ (m:n = 1:5) , and $Bi_4Te_3$ (m:n = 3:3), each exhibiting distinct physical properties.[18, 19] Notably, layered superstructures derived from alternating Bi double layers and $Bi_2Te_3$ blocks, such as $(Bi_2)_1(Bi_2Te_3)_1$ (i.e., $Bi_4Te_3$), exhibit enhanced carrier densities and modified band structures compared to the individual Bi layers or $Bi_2Te_3$ blocks. These effects arise from strong interlayer interactions and electron injection form the $Bi_2$ layer into the $Bi_2Te_3$ layer.[20] Such structural adaptability, coupled with the intrinsic anisotropy of the Bi-Te system, emphasizes the importance of compositional and stacking control in tailoring the thermoelectric and quantum functionalities of these materials.[18-20, 23]

The structural diversity and fascinating properties of the Bi-Te system have motivated extensive research; however, the controlled colloidal synthesis of well-defined nanostructures across different phases remains relatively unexplored. In this work, we developed a phosphine-free colloidal approach that enables the synthesis of large $Bi_2Te_3$ nanosheets with lateral sizes up to 10 μm. By simply tuning the reaction temperature, we also achieved selective synthesis of BiTe and $Bi_4Te_5$. The phase transformation behaviors among these phases were systematically studied using temperature-dependent X-ray diffraction (XRD), *in situ* transmission electron microscopy (*in situ* TEM), and complementary techniques, offering deeper insights into the kinetic and thermodynamic factors that govern phase stability and conversion in the Bi-Te system at the nanoscale. This study deepens the fundamental understanding of phase relationships in bismuth tellurides, offering valuable insights into their evolutionary pathways and providing a useful reference for exploring their potential in future thermoelectric, spintronic, and topological systems.

## RESULTS AND DISCUSSION

$Bi_2Te_3$ nanosheets were synthesized by hot-injection of a tellurium precursor into a solution containing bismuth acetate in oleic acid and diphenyl ether. In the conventional route, 0.2 mmol of TOP-Te was injected at 170 °C. For the phosphine-free route, diphenyl ditelluride in oleylamine (DPD-Te, 0.2 mmol) was injected at 250 °C, with all other conditions kept identical. The resulting nanosheets were purified by repeated precipitation and centrifugation. Full synthetic details are provided in the Experimental Section.

When TOP-Te (tri-octylphosphine telluride) was used as the Te precursor, the product was crystallized in the form of $Bi_2Te_3$ nanosheets. They were analyzed using transmission electron microscopy (TEM). Figure 1 shows their TEM images at different magnifications, revealing that the well-defined hexagonal sheet-like crystals can reach lateral dimensions of up to 2 μm (Figure 1c). Figure 1d shows the selected area electron diffraction (SAED) pattern of a representative $Bi_2Te_3$ nanosheet, displaying six well-defined diffraction spots along the [001] zone axis. The hexagonal arrangement of the well-resolved single diffraction spots, each corresponding to a specific Miller index from the {*hk0*} set, confirms that the nanosheet is single-crystalline. As shown in Figure 1e, the X-ray diffraction (XRD) pattern of the synthesized nanosheets is in good agreement with the standard pattern of $Bi_2Te_3$ (ICDD PDF card No. 01-087-7710). All detected peaks can be indexed to the rhombohedral $Bi_2Te_3$ (space group R-3m, No. 166), with the structure described by a hexagonal cell containing three formula units.

Interestingly, our previous SnTe nanosheet synthesis showed that the injection of 1 mmol TOP before the TOP-Te precursor was crucial for stabilizing two-dimensional growth.[17] In the present

Bi-Te system, however, the same procedure led to a completely different outcome. When 1 mmol of TOP was injected before the addition of the TOP-Te precursor, the reaction mixture rapidly turned from clear to semi-transparent with noticeable scattering. We therefore terminated the reaction without injecting the TOP-Te precursor. Characterization of the product (Figure S1, Supporting Information) revealed the formation of metallic Bi nanosheets with lateral sizes of approximately 2 μm, consistent with our previous work on Bi nanosheet synthesis.[24] This indicates that, in the Bi system, TOP primarily acts as a strong reducing agent that reduces $Bi^{3+}$ to metallic Bi,[17, 25] rather than functioning as a capping ligand to promote two-dimensional growth as observed in the SnTe case.[17] The XRD pattern of the obtained product matches well with the standard pattern of Bi (COD card No. 96-231-0890), confirming that the crystalline phase is predominantly metallic Bi (Figure S1b). Moreover, energy-dispersive X-ray spectroscopy (EDS) analysis and elemental mapping further demonstrate that the nanosheets are mainly composed of Bi, with only minor amounts of Te dispersed on the sheet surfaces (Figure S1c-f). These results indicate that the introduction of TOP, rather than stabilizing the two-dimensional $Bi_2Te_3$ structure, preferentially promotes the rapid reduction of $Bi^{3+}$ to $Bi^0$, leading to the formation of metallic Bi nanosheets (as shown in Figure S1 and schematically illustrated in Figure 2a). In addition, TOP also acts as a strong capping ligand that critically influences the nucleation and growth kinetics through an "arrested growth mechanism".[25, 26] Specifically, due to its high binding strength to metal surfaces, TOP rapidly coordinates with both the precursor species and the growing Bi nuclei. This surface passivation prevents further attachment of Bi monomers, thereby "arresting" the lateral growth at an early stage even when unreacted precursors remain in solution. This phenomenon commonly referred to in the literature as the "arrested growth mechanism".[25, 26] As a result, nanoparticle or nanosheet growth is terminated at an early stage, even in the presence of excess unreacted precursor.[26, 27] While this mechanism is beneficial for achieving narrow size distributions, it inherently limits the lateral growth of Bi nanosheets, rendering the formation of larger structures difficult through simple optimization of reaction parameters. Consequently, to overcome this intrinsic growth limitation caused by strong ligand passivation, we developed a "TOP-free" colloidal synthesis approach using diphenyl ditelluride (DPD-Te) as the tellurium precursor.

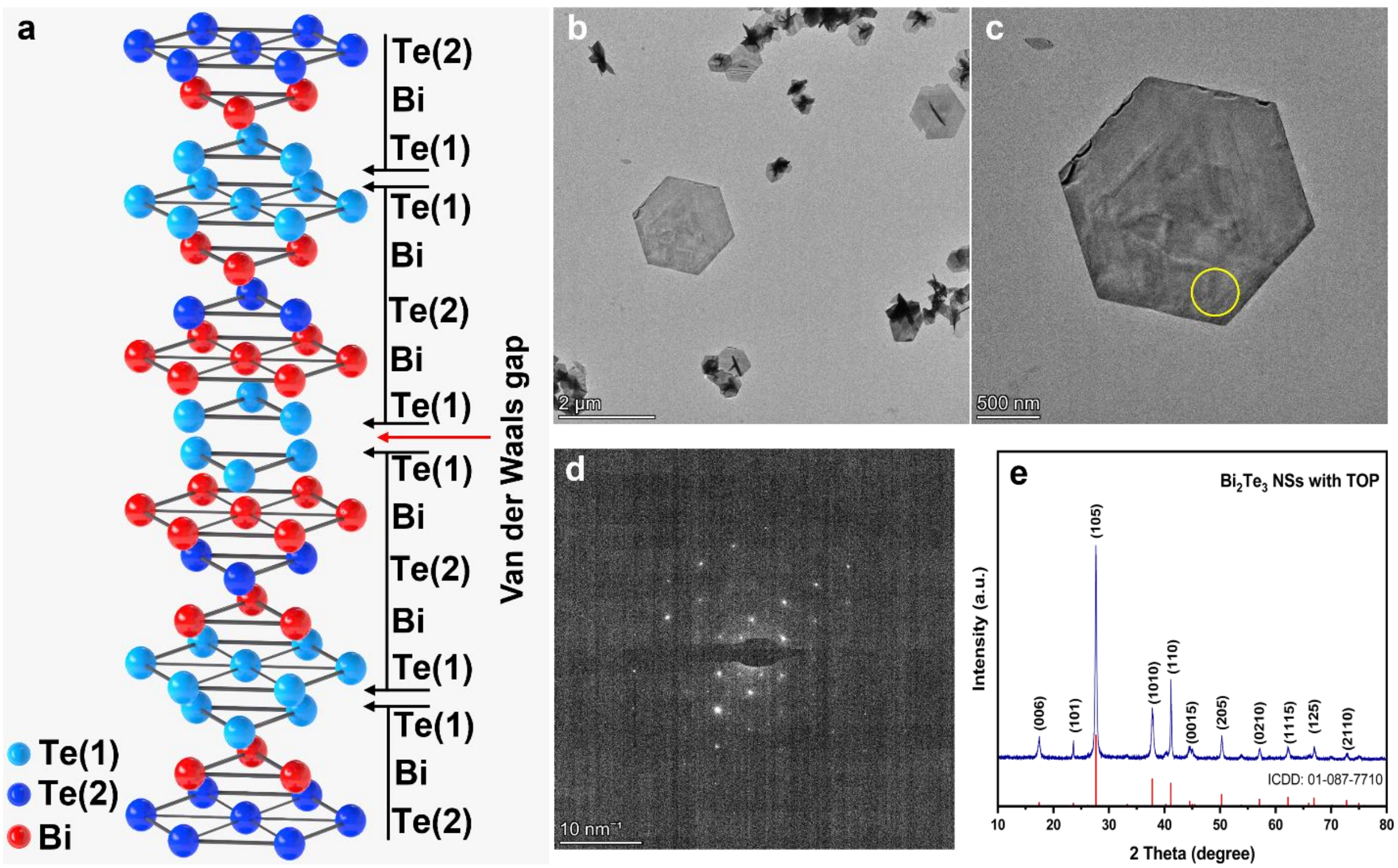

**Figure 1.** (a) Schematic crystal structure of $Bi_2Te_3$, highlighting the quintuple-layer stacking sequence (Te–Bi–Te–Bi–Te). TEM images at different magnifications: (b) showing the overall distribution of hexagonal $Bi_2Te_3$ nanoplatelets and (c) an individual nanosheet with a well-defined hexagonal shape. (d) Corresponding SAED pattern obtained from the nanoplatelet shown in (c), displaying six-fold symmetric diffraction spots. (e) XRD pattern of $Bi_2Te_3$ nanoplatelets, consistent with the hexagonal $Bi_2Te_3$ crystal structure (ICDD PDF card No. 01-087-7710).

After replacing the TOP-Te precursor with DPD-Te (diphenyl ditelluride),[28] the synthesized $Bi_2Te_3$ nanosheets exhibit significantly increased lateral dimensions and more uniform hexagonal morphologies. While nanosheets obtained from TOP-Te precursors reached a maximum size of only 2 μm, those synthesized with DPD-Te display an average lateral size of approximately 5.2 μm, with the largest nanosheets extending up to 10 μm (Figure S2). A representative TEM image (Figure 2b) reveals a well-defined hexagonal nanosheet, and the corresponding SAED pattern confirms its single-crystalline nature with hexagonal symmetry. Scanning transmission electron microscopy (STEM) imaging reveals the atomically resolved lattice with the electron beam along the [001] direction. An interplanar spacing of 0.22 nm, corresponding to the (110) plane of hexagonal $Bi_2Te_3$ is obtained (Figure S3). Atomic force microscopy (AFM) analysis indicates a nanosheet thickness of approximately 23 nm (Figure 2c-d). The XRD pattern of the sample (Figure 2e) is in agreement with the standard ICDD PDF card (No. 00-015-0863), and the sharp, intense peaks indicate the high crystallinity and phase purity of the synthesized $Bi_2Te_3$ nanosheets.

To gain deeper insight into the reactivity difference between the two precursors, we performed density functional theory (DFT) calculations (see Experimental section for details).[29, 30] As summarized in Table S1 (Supporting Information), the TOP-Te adduct exhibits a dissociation energy of 75.2 kJ/mol, indicating higher thermodynamic stability relative to the dissociated TOP + Te fragments. In contrast, the dissociation of diphenyl ditelluride into two phenyl-Te radicals requires only 39.6 kJ/mol, and the related PhTe• → Ph• + Te process has an energy of 93.4 kJ/mol. These results suggest that, although both precursors are stable at room temperature (kT = 2.48 kJ/mol), the DPD-Te precursor likely undergoes a stepwise decomposition. The initial facile formation of phenyl-Te radicals is followed by a higher-energy step for free Te release (93.4 kJ/mol), which facilitates the formation of a lesser number of nuclei in comparison to TOP-Te and provides a milder and more controlled delivery of reactive Te species compared to the direct dissociation from the TOP-Te adduct. This slower Te-supply rate effectively suppresses secondary nucleation and promotes the lateral growth of $Bi_2Te_3$ nanosheets, accounting for the larger lateral dimensions and superior crystallinity observed in the phosphine-free DPD-Te synthetic route.

In addition to the precursor effect, once the Te-precursor becomes DPD-Te, the reaction temperature plays a decisive role in determining the phase formation. At 230 °C or below, the products consist predominantly of BiTe nanosheets (Figure 2f-g and Figure S4), whereas at 240 °C, $Bi_4Te_5$ nanosheets are obtained (Figure 2h-i). Phase-pure $Bi_2Te_3$ nanosheets are successfully synthesized at 250 °C. Both BiTe and $Bi_4Te_5$ belong to the Bi–Te homologous series of layered $(Bi_2)_m(Bi_2Te_3)_n$, which are derived from different stacking sequences of hexagonal Bi double layers and $Bi_2Te_3$ blocks (as schematically shown in Figure 2a).[18, 19] As already mentioned, the "end-members" of this series are metallic Bi and semiconducting $Bi_2Te_3$, and the intermediate members, such as BiTe and $Bi_4Te_5$, belong to infinitely adaptive series that can be described using a modulated structural approach. To further verify the crystallinity and structural characteristics of these intermediate phases, TEM and STEM analysis were performed. As evidenced by the TEM image and the corresponding SAED pattern in Figure S5, the BiTe and $Bi_4Te_5$ nanosheets display similar six well-defined diffraction spots along the [001] zone axis, confirming their single-crystalline nature and hexagonal symmetry. In addition, high-resolution STEM imaging clearly reveals lattice fringes with interplanar spacings of ~0.21 nm and ~0.22 nm, corresponding to the (110) planes of hexagonal BiTe and $Bi_4Te_5$, respectively (Figure S6 and later in Figure 4c).

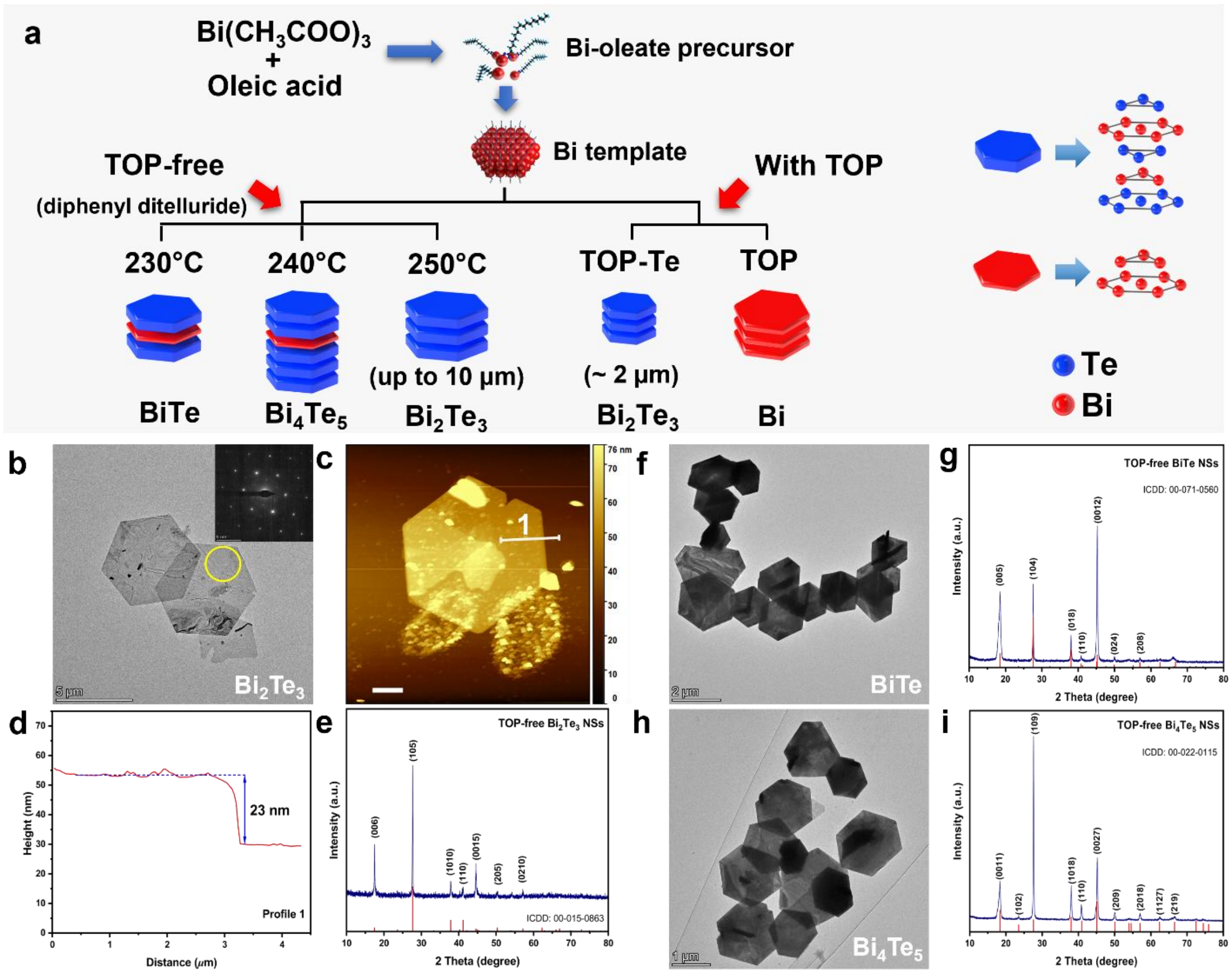

**Figure 2.** (a) Schematic illustration of the morphological evolution and phase formation of Bi–Te NSs during colloidal synthesis. (b) TEM image of $Bi_2Te_3$ NSs with the corresponding SAED pattern inset under the "TOP-free" approach. (c, d) AFM image (scale bar = 2 μm) and thickness profile of $Bi_2Te_3$ NSs. (e) XRD pattern of $Bi_2Te_3$ NSs, consistent with the hexagonal $Bi_2Te_3$ crystal structure (ICDD card No. 00-015-0863). (f) TEM image and (g) XRD pattern of BiTe nanoplatelets, consistent with the hexagonal BiTe crystal structure (ICDD card No. 00-071-0560). (h) TEM image and (i) XRD pattern of $Bi_4Te_5$ nanoplatelets, consistent with the hexagonal $Bi_4Te_5$ crystal structure (ICDD card No. 00-022-0115).

Figure 3a exhibits a high-angle annular dark-field scanning transmission electron microscopy (HAADF-STEM) image of a representative hexagonal $Bi_2Te_3$ nanosheet. This method is highly effective for studying the local structure and composition of materials at the atomic level and can be combined with electron energy loss spectroscopy (EELS). An EELS map was acquired over the nanosheet, and spectra were extracted from two representative regions – the center and the edge of the nanosheet – as shown in Figure 3a. In our previous studies on colloidally synthesized Bi nanosheets, we observed that the Bi nanosheet edges are oxidized, leading to pronounced changes in the EELS spectra.[24] Motivated by this observation, EELS spectra from the center (Figure 3b) and edge (Figure 3c) regions of the $Bi_2Te_3$ nanosheet were compared in this study. The highly similar spectral features obtained from both positions indicate the absence of oxidation, suggesting good chemical stability of the $Bi_2Te_3$ nanosheets. The EELS spectra show distinct features at 5.6 eV and 16.3 eV energy losses, with the latter corresponding to the bulk plasmon peak, which is consistent with previous reports.[31] Additionally, peaks observed at 25.8 eV and 27.9 eV are attributed to electronic transitions from the Bi $5d_{5/2}$ and Bi $5d_{3/2}$ core levels to the conduction band minimum, in good agreement with literature findings.[31, 32] Notably, the feature at 5.6 eV is identified as a surface plasmon mode intrinsic to $Bi_2Te_3$, consistent with previous reports.[33] EELS spectra obtained for both BiTe and $Bi_4Te_5$ nanoplatelets reveal almost identical spectral features (Figure S8b and S9b), suggesting that their electronic structures and local bonding environments are very similar. EDS analysis confirms that Bi and Te are homogeneously distributed throughout the surface of the $Bi_2Te_3$ nanosheets (Figure 3d-f). The corresponding elemental mapping clearly illustrates a uniform spatial distribution of both elements without noticeable segregation or clustering. Furthermore, the quantitative analysis indicates that the atomic ratio of Bi to Te is approximately 1:1.5 (Figure S7), which closely matches the theoretical stoichiometric composition of $Bi_2Te_3$. This consistency between the experimental data and the expected chemical formula provides strong evidence for the successful synthesis of phase-pure $Bi_2Te_3$ nanosheets. Similarly, the EDS analysis of BiTe and $Bi_4Te_5$ nanosheets reveals consistent results. The elemental mapping demonstrates that Bi and Te are uniformly distributed across the sheets (Figure 3g-n). For BiTe, the Bi:Te atomic ratio is close to 1:1 (Figure S8c), while for $Bi_4Te_5$ it is about 1:1.2 (Figure S9c). Both values are highly consistent with their respective stoichiometric compositions, thereby further confirming the optimal conditions of the synthesis process and the phase purity of the obtained nanosheets.

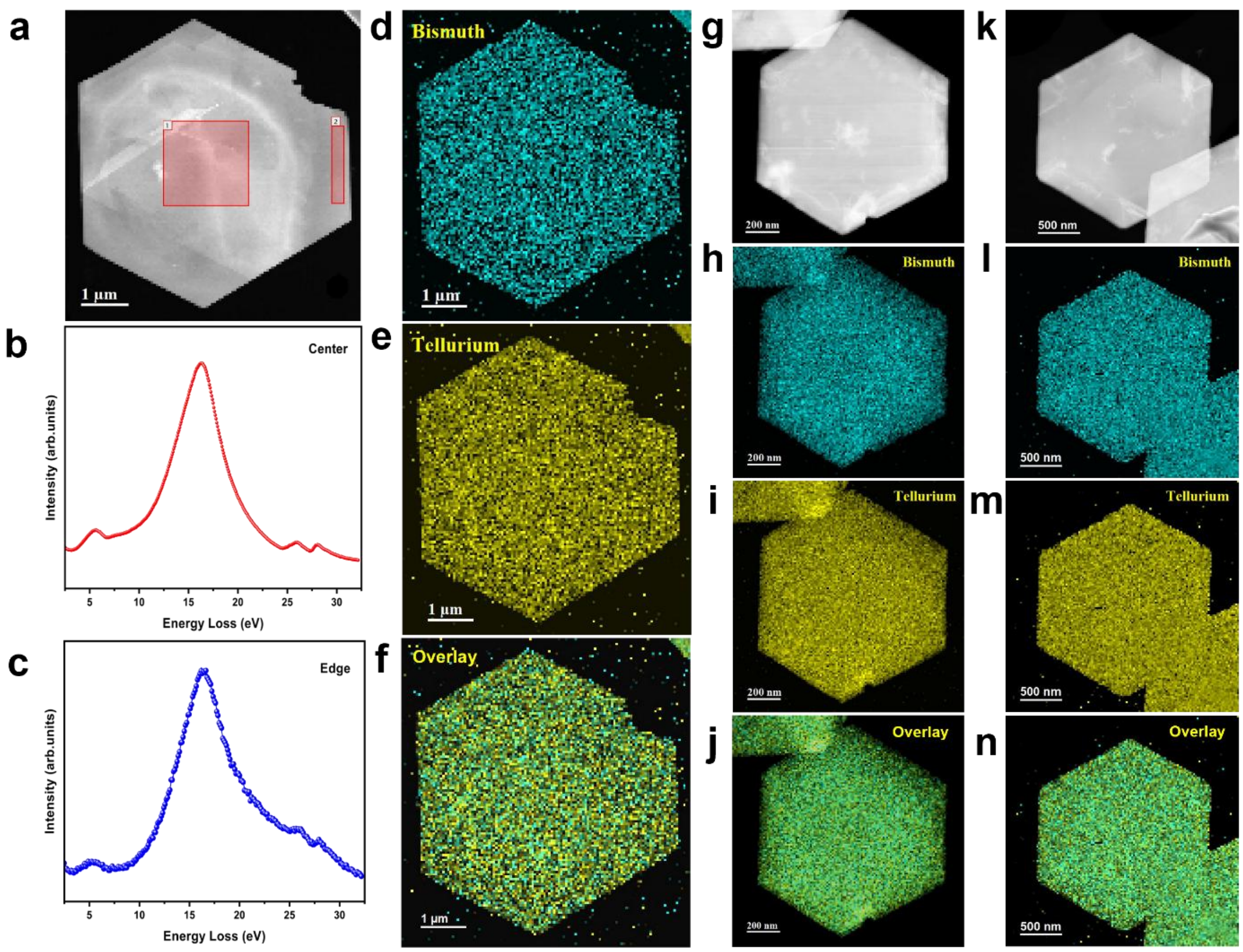

**Figure 3.** (a) HAADF-STEM image of an individual $Bi_2Te_3$ nanosheet; the red square marks the area analyzed by EELS. (b, c) EELS spectra extracted from the selected region at the (b) center

and (c) edge of the $Bi_2Te_3$ nanosheet. Corresponding EDS elemental maps: (d) Bi, (e) Te, and (f) overlay map of Bi and Te. (g) HAADF-STEM image of BiTe nanosheet with corresponding EDS maps: (h) Bi, (i) Te, and (j) overlay map of Bi and Te. (k) HAADF-STEM image of $Bi_4Te_5$ nanosheet with corresponding EDS maps: (l) Bi, (m) Te, and (n) overlay map of Bi and Te.

To better elucidate the structural relationships and temperature-induced phase transformation among BiTe, $Bi_4Te_5$, and $Bi_2Te_3$ nanosheets, we conducted temperature-dependent XRD analyses in air. The XRD patterns of BiTe and $Bi_4Te_5$ nanosheets were recorded at incremental temperatures to monitor the change of crystal structures (Figure S10a and S10c). As shown in Figure 4a, the diffraction profiles of BiTe nanosheets exhibit pronounced temperature-dependent changes. At lower temperatures (below ca. 330 °C), the diffraction pattern is dominated by the reflections of the initial phase, with intense peaks at $2\theta \approx 17.8°$, which can be indexed to the (005) planes of the BiTe crystalline phase (Figure 4a and Figure S10a). Upon heating, these peaks gradually decrease in intensity and eventually vanish, while new reflections emerge at $2\theta \approx 17.3°$ and 8.4° (after heating at 340°C), corresponding to the (006) and (003) planes of $Bi_2Te_3$. As the temperature further increases, these $Bi_2Te_3$ peaks vanish near 400 °C, giving rise to a new peak at $2\theta \approx 7.4°$, which likely comes from the (001) plane of $Bi_2TeO_5$. The phase transformation of $Bi_4Te_5$ is particularly well resolved: it transforms into $Bi_2Te_3$ at around 330 °C, and completely converts to $Bi_2TeO_5$ at 360 °C (Figure 4b and Figure S10c). To quantify this transformation, we performed peak fitting on the peaks marked with different colors in Figure 4a-b, and plotted their $2\theta$ positions as a function of temperature. The resulting diagram clearly illustrates the sequential transformation of the crystalline phases (Figure S10b and 10d).

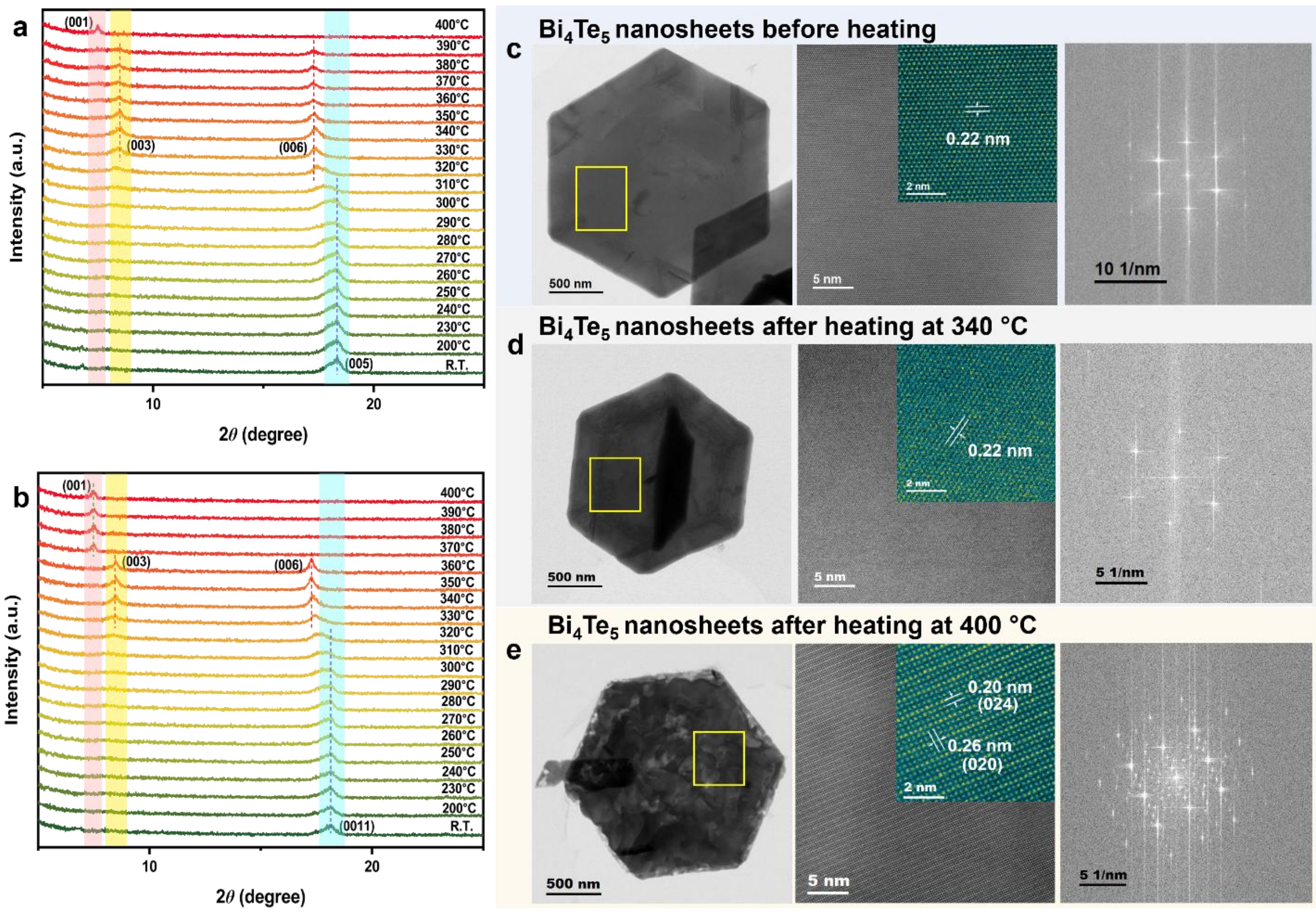

**Figure 4.** (a, b) Selected regions of the XRD patterns of BiTe (a) and $Bi_4Te_5$ (b) nanosheets recorded over the temperature range from room temperature to 400 °C. To quantitatively analyze the temperature-dependent structural transformation, peak fitting was performed on the diffraction peaks indicated by different colors. (c) STEM image of $Bi_4Te_5$ nanosheets before heating; corresponding HAADF-STEM image and the associated FFT pattern. (d) STEM image of $Bi_4Te_5$ nanosheets after heating at 340 °C; corresponding HAADF-STEM image and the associated FFT pattern. (e) STEM image of $Bi_4Te_5$ nanosheets after heating at 400 °C; corresponding HAADF-STEM image with a magnified false-color inset for clarity, and the associated FFT pattern.

The reproducibility of the results in two independent XRD measurements (Figure 4a-b) confirms the reliability of the observed temperature-induced phase transformation. These investigations demonstrate that the material undergoes a distinct crystalline-to-crystalline phase transformation upon heating, driven by atomic-scale rearrangements and local compositional redistribution rather than a classical lattice reconstruction or polymorphic transition. According to the most recent thermodynamic assessment of the Bi-Te binary system,[34, 35] $Bi_2Te_3$ represents the thermodynamically most stable phase in the Te-rich region. In contrast, BiTe is a peritectic phase that forms via the invariant peritectic reaction Liquid + $Bi_2Te_3$ → BiTe at a fixed temperature, whereas $Bi_4Te_5$ is a metastable compound.[34, 35] Both phases also exhibit a extending toward the Bi-rich side, which is accommodated by anti-site defects in BiTe (Bi atoms occupying Te sites) and antistructure defects of tellurium atoms in $Bi_4Te_5$, giving rise to additional bismuth layers.[35] The crystallization of these Bi-rich phases at relatively low temperatures (230 °C for BiTe NSs and 240 °C for $Bi_4Te_5$ NSs) is attributed to kinetic trapping enabled by rapid nucleation and strong undercooling effects in the Bi-Te system.[35] Importantly, the relative thermal stability observed in our nanosheets follows the same stability trend as predicted by the bulk Bi-Te phase diagram.

To gain deeper insights into the structural transformation mechanisms, *ex situ* heating experiments were performed on BiTe- and $Bi_4Te_5$-containing nanosheets directly supported on Ni grids under ambient air, followed by detailed HRTEM and EDS characterizations (Figure 4c-e and Figure S11 to S14). Nickel grids were chosen because, compared to copper (melting point 1084.6°C), nickel has a higher melting point (1453 °C) and better thermal stability, minimizing deformation or oxidation of the supporting carbon film during heating. Indeed, when copper grids were used, the carbon film exhibited severe wrinkling and damage at 340 °C.

When BiTe nanosheets were heated to 340 °C, they largely retained their morphology, and EDS elemental maps showed a uniform distribution of Bi and Te, with only trace amounts of oxygen (Figure S11). Correspondingly, the FFT patterns of the HAADF-STEM images displayed a clear hexagonal symmetry with an interplanar spacing of 0.22 nm, which matches well with the (110) plane of $Bi_2Te_3$ and confirms the presence of a hexagonal layered structure after the transformation (Figure S12). $Bi_4Te_5$ exhibits the same behavior after heating to 340 °C (Figure 4d and Figure S13). Upon further heating to 400 °C, a distinct structural change occurred. Although the EDS mappings still revealed a homogeneous distribution of Bi and Te, a significant oxygen signal emerged, indicating oxidation of the $Bi_4Te_5$ nanosheets (Figure S14). The FFT patterns of the HAADF-STEM images at this temperature exhibited an orthorhombic symmetry with lattice spacings of 0.20 nm and 0.26 nm (Figure 4e), corresponding to the (024) and (020) planes of $Bi_2TeO_5$, respectively. These microstructural observations are in agreement with the temperature-dependent XRD results, confirming that BiTe and $Bi_4Te_5$ first transform into $Bi_2Te_3$ around 340 °C and subsequently oxidize completely to $Bi_2TeO_5$ at 400 °C. Furthermore, the EELS spectra corroborate these findings, reinforcing the sequential crystalline-to-crystalline transition and further oxidation process of the nanosheets upon heating (Figure S11 and Figure S13 to S14).

In addition, the thermal stability of BiTe and $Bi_4Te_5$ nanosheets were investigated using *in situ* TEM under high-vacuum conditions. We find that the transformation pathways differ from those at ambient pressure. Based on the above observations, the temperature-dependent XRD measurements performed under ambient atmospheric conditions confirm that both BiTe and $Bi_4Te_5$ transform into $Bi_2Te_3$ upon heating to approximately 340 °C. Considering the relatively low melting point of bulk $Bi_2Te_3$ (586 °C)[36] and the partial pressures of Te (~$10^{-1}$Pa) and Bi (~$10^{-6}$ Pa) at 350–500 °C under the applied vacuum ($10^{-5}$ Pa),[36] *in situ* TEM heating is expected to promote sublimation under such conditions. Figure 5 illustrates the temperature-dependent structural changes of BiTe nanosheets. When the temperature increased from 25°C to 340°C, the BiTe nanosheet remained structurally stable. EDS measurements reveal no significant change in Bi and Te content during heating, with a Bi:Te ratio consistently around 1:1.1, confirming the compositional stability of BiTe (Figure S15). As the temperature approaches 400°C, BiTe gradually sublimates from the edges, with the process initiating at the corners highlighted by the yellow dashed lines in Figure S16c, ultimately consuming the entire thickness of the nanosheets. At this stage, the Te content drops sharply (Figure S17a), and sublimation accelerates. In the residual portions of the nanosheets (Figure 5f), EDS reveals only a small amount of Te, with the composition being predominantly Bi (Figure S15d).

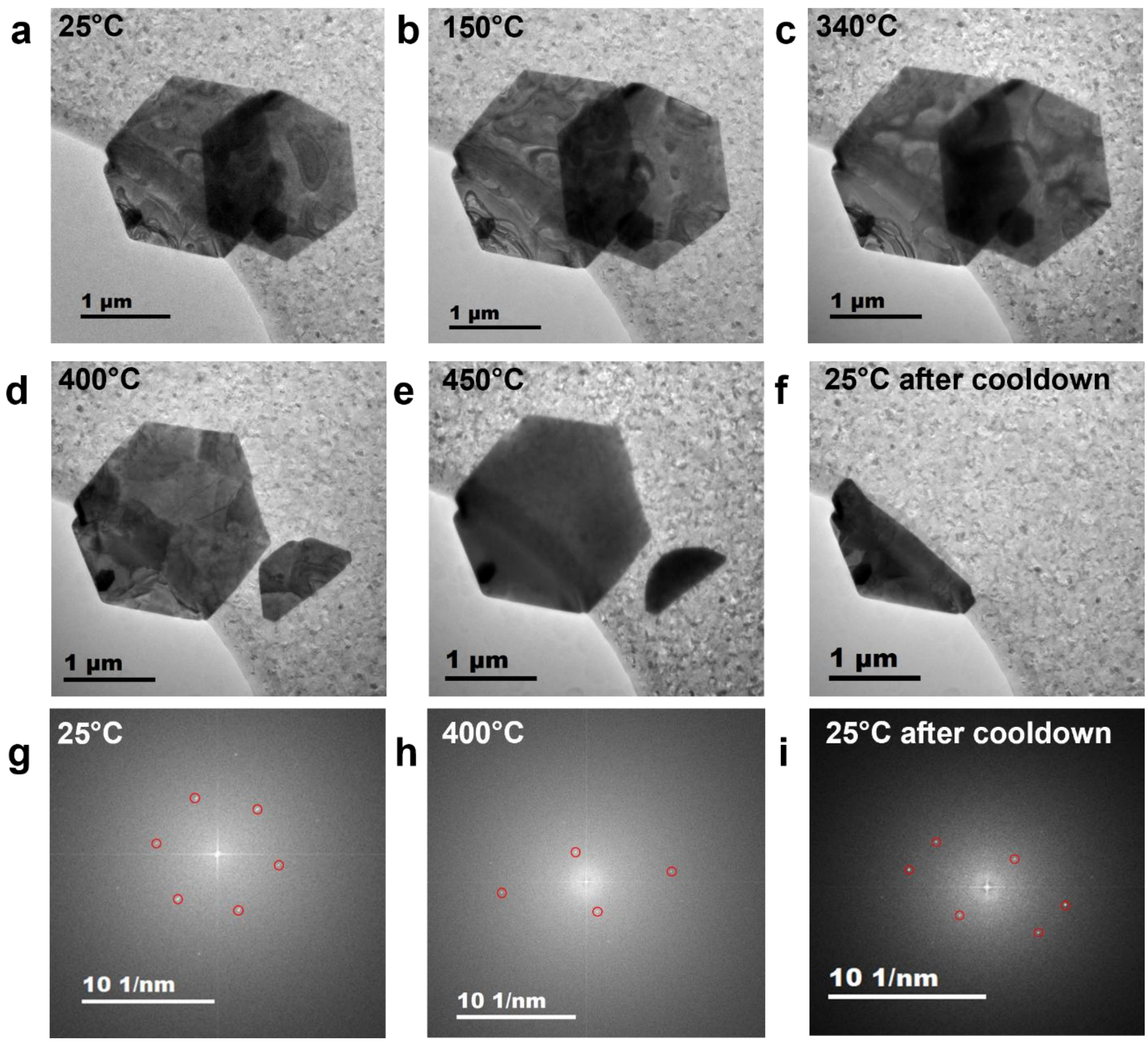

**Figure 5.** (a-f) *In situ* TEM images of BiTe nanosheets recorded during heating from 25 °C to 450 °C and subsequent cooling. The bright background contrast originates from a hole (≈8 μm in diameter) of the ceramic heater covered with a SiN film. Due to limitations in sample preparation, the nanosheet could not be positioned completely within the hole and is partially supported at the hole edge throughout the experiment. (g-i) FFT patterns from the center of the left nanosheet at different temperatures, showing the changes of the crystal structure during heating (key features are highlighted with red circles for clarity). The images correspond to 25 °C (g), 400 °C (h), and after heating to 450 °C followed by cooling back to 25 °C (i).

The mechanism of BiTe sublimation is illustrated in more detail in Video S1, which captures the progressive sublimation from a corner of a BiTe nanosheet at 400 °C. Selected frames from the video, presented in Figure S16, reveal that sublimation initially proceeds through the stepwise decomposition of atoms from the nanosheet edge with progression of facets corresponding to the outer faces of the nanocrystal. At a later stage, sublimation accelerates, and irregular edge forms are clearly visible in Figure S16, where they are marked by yellow dashed lines and red circles. These results indicate not only complete sublimation of portions of the nanosheets but also a strong preferential loss of Te before Bi and prior to visible structural disintegration. This behavior is consistent with previous studies on the phase transformation of $Bi_2Te_3$,[36] upon heating which attribute the preferential sublimation of Te to the lower activation energy for its evaporation (52.5 kJ $mol^{-1}$) compared with that of Bi (104.8 kJ $mol^{-1}$).[37] Figure 5g-i present the corresponding FFT patterns recorded before heating, at 400 °C, and after heating to 450 °C followed by cooling to 25 °C. The systematic structural changes of the diffraction spots across these three stages provides direct crystallographic evidence for the phase transformation occurring during the entire heating cycle.

The thermal stability of $Bi_4Te_5$ nanosheets follows a trend similar to that of BiTe. Figure S18 shows the morphology of $Bi_4Te_5$ nanosheets at different temperatures. The nanosheets remain structurally stable upon heating to 340 °C; EDS analysis, however, reveals a slight increase in Te content, with the Bi:Te ratio shifting from ~1:2 to ~1:1.25, reflecting the gradual trend of $Bi_4Te_5$ transforming toward $Bi_2Te_3$ (Figure S19). Notably, consistent with the temperature-dependent XRD results, the nanosheets begin to decompose from the edges at ~360 °C (Figure S18d). The detailed sublimation mechanism is captured in Video S2, which shows stepwise decomposition from the nanosheet edges. Furthermore, the temperature-dependent Bi and Te contents extracted from EDS, presented in Figure S20, confirm that the sublimation of $Bi_4Te_5$ nanosheets follows a behavior similar to that of BiTe, with Te preferentially lost before Bi and prior to any obvious structural disintegration. The corresponding FFT patterns recorded at different heating temperatures (Figure S18i-l) reveal a systematic change of the diffraction spots across the three stages, providing direct crystallographic evidence for the structural changes with increasing temperature.

Our experiments indicate that BiTe and $Bi_4Te_5$ exhibit lower structural stability than $Bi_2Te_3$. Upon heating, both phases tend to transform into $Bi_2Te_3$ under air or vacuum. In air, the transformed $Bi_2Te_3$ is eventually oxidized, whereas under high vacuum, preferential Te sublimation leads to progressive nanosheet decomposition. Based on our observations and previous reports, we propose the following phase transformation mechanism. Both BiTe and $Bi_4Te_5$ can be described as layered compounds of the general formula $(Bi_2)_m(Bi_2Te_3)_n$ (where $m$ and $n$ are integers), featuring intrinsic $Bi_2$ bilayers interleaved with $Bi_2Te_3$ quintuple layers. To illustrate the underlying transformation mechanism, we focus on BiTe as a representative example. BiTe is a natural van der Waals-like heterostructure, with a $Bi_2$ bilayer sandwiched between two [Te-Bi-Te-Bi-Te] quintuple layers.[23, 38] X-ray pair distribution function analysis[23] indicates that the distance between the two terminal Te atoms (Figure 1a) in consecutive quintuple layers in $Bi_2Te_3$ is 3.6 Å. In contrast, the distance between the Bi (in the $Bi_2$ bilayer) and the terminal Te of the quintuple layer in BiTe is 3.47 Å. Both distances are smaller than the sum of the van der Waals radii of Bi (2.3 Å) and Te (2.1 Å) and larger than the sum of the ionic radii ($Bi^{3+}$= 1.03 Å; $Te^{2-}$ = 1.40 Å). The Bi-Te bond length within the quintuple layer in BiTe (3.09 Å) is slightly longer than in $Bi_2Te_3$ (3.04 Å), indicating a weakening of bonds and a comparatively softer lattice.[23] This underpins the lower structural stability of BiTe compared to $Bi_2Te_3$, thereby rendering BiTe thermodynamically predisposed to transform into the more stable $Bi_2Te_3$ phase.

These structural features, characterized by a comparatively softer lattice and weakened Bi-Te bonds, suggest that the central $Bi_2$ bilayers are the most thermally labile components. This structural instability is further highlighted by the distinct morphology of the as-synthesized nanosheets. AFM measurements indicate that $Bi_2Te_3$ nanosheets have a thickness of approximately 23 nm (Figure 2c-d), while EELS relative thickness reveals that BiTe ($t/\lambda$ = 1.31, Figure S8) and $Bi_4Te_5$ ($t/\lambda$ = 1.03, Figure S9) nanosheets are significantly thicker, approximately 5 times that of $Bi_2Te_3$ ($t/\lambda$ = 0.24, Figure S7). This comparison of relative thickness ($t/\lambda$) assumes that the inelastic mean free path $\lambda$ remains relatively constant across these chemically and structurally related Bi-Te phases. This increased thickness is intrinsically linked to the $(Bi_2)_m(Bi_2Te_3)_n$ homologous structure, where the interleaving of multiple $Bi_2$ bilayers promotes vertical growth.[39] Crucially, despite being nearly 5 times thicker, BiTe and $Bi_4Te_5$ exhibit lower thermal stability than the thinner $Bi_2Te_3$ sheets. This observation suggests that the phase transformation is primarily governed by internal structural chemistry rather than simple size effects. According to the Bi-Te phase diagram, the $Bi_2$ bilayers are expected to exhibit significantly enhanced atomic mobility or "quasi-liquid" behavior at temperatures exceeding 270°C (near the Bi melting point of 271.4°C). In an ambient atmosphere, this thermal trigger drives the diffusion of Bi atoms from the internal bilayers toward the $Bi_2Te_3$ quintuple layers and the surface (Figure 6). This migration process is supported by our XRD analysis (Figure 4a). At 280°C, the broadening of the BiTe (005) reflection and subsequent Scherrer analysis indicate a progressive reduction in the crystalline domain thickness of the BiTe phase (Figure S10e). As Bi atoms reach the surface and undergo slow oxidation to form an amorphous layer, the structure reorganizes with the $Bi_2Te_3$ quintuple layers becoming the dominant remaining component. By 320°C, the $Bi_2Te_3$ phase begins to stabilize, evidenced by the emergence of $Bi_2Te_3$ (003) and (006) peaks (Figure 4a). Scherrer analysis of the (006) reflection shows that the $Bi_2Te_3$ domain thickness grows as the transformation completes, reaching a maximum at 340°C (Figure S10e). Beyond this temperature, the $Bi_2Te_3$ thickness begins to decrease due to continued surface oxidation. This amorphous oxide layer eventually crystallizes into $Bi_2TeO_5$ at 400°C, at which point it becomes detectable by XRD.

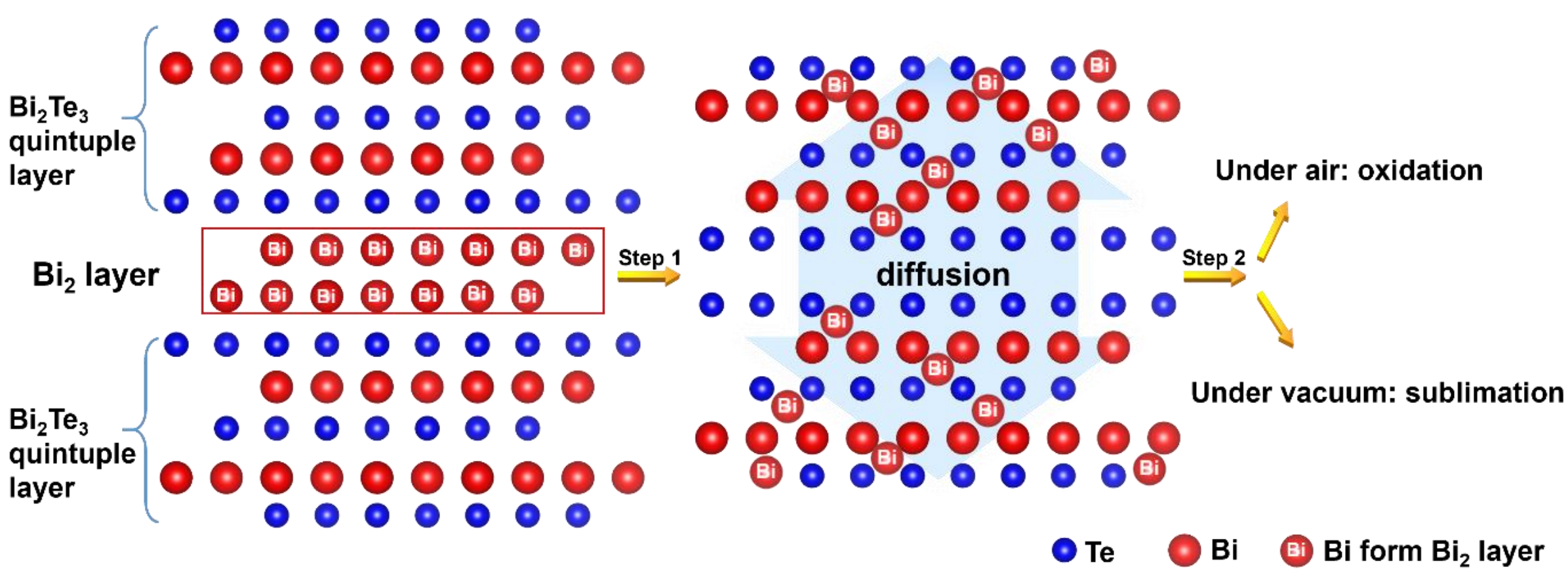

**Figure 6.** Schematic illustration of the phase transformation mechanism of BiTe nanosheets.

During *in situ* TEM measurements under vacuum, a similar destabilization of the $Bi_2$ bilayer occurs, yet the absence of oxygen fundamentally alters the compositional evolution. Upon heating, the rapid diffusion of the $Bi_2$ layers causes a noticeable reduction in nanosheet thickness, as directly visualized in Video S1. However, unlike the air-based process, the diffused Bi atoms remain trapped on the surface or within the lattice rather than being consumed by oxidation. Consequently,

EDS measurements show that the Bi:Te atomic ratio remains largely constant during this initial restructuring phase. The nanosheet maintains its integrity until temperatures exceed 340°C, above which the structural evolution becomes dominated by the preferential sublimation of Te. Furthermore, the in-situ TEM imaging (Figure 5 and Figure S16) also reveals slight changes in surface morphology and thickness across different heating stages. Above 340 °C, the structural evolution is predominantly governed by the rapid evaporation of Te. The lower activation energy for Te evaporation (52.5 kJ $mol^{-1}$ vs. 104.8 kJ $mol^{-1}$ for Bi)[37] drives preferential Te sublimation. This interpretation is corroborated by EDX measurements, which reveal a progressive decrease in the Bi:Te ratio from ~1:1.1 to ~1:0.63 and ultimately to ~1:0.14, highlighting the accelerated Te loss and consequent compositional evolution at increased temperatures.

These observations, together with previously reported notable properties of the $Bi_2$ bilayer in bismuth-telluride compounds,[18, 22, 23, 38, 40] serve as further motivation for future investigations. Specifically, the structural versatility of the $(Bi_2)_m(Bi_2Te_3)_n$ series, and the ability to drive phase transitions between them, highlights their potential as a tunable platform for prospective applications. While stoichiometric $Bi_2Te_3$ remains the most robust and widely utilized phase for high-performance thermoelectrics and topological studies, the introduction or removal of $Bi_2$ bilayers offers a potent strategy for material property tuning.[19] Literature suggests that for thermoelectrics, $Bi_2$ blocks may potentially act as strong electron donors that can optimize the power factor through interfacial energy filtering effects.[18, 20] For quantum applications, the enhanced metavalent bonding in Bi-rich phases (e.g., BiTe) facilitates electronic topological transitions (ETT) at lower pressures compared to $Bi_2Te_3$.[23] Thus, the observed thermal and structural evolution in these nanosheets not only highlights their complex chemistry but also provides useful insights that may guide the future development of Bi-Te materials for potential electronic or spintronic systems.

## CONCLUSION

In summary, we have developed a "TOP-free" colloidal synthesis approach for producing $Bi_2Te_3$, BiTe, and $Bi_4Te_5$ nanosheets with well-controlled morphology and phase purity. By carefully tuning the reaction temperature and precursor chemistry, phase-selective formation of these layered bismuth-telluride compounds was achieved, highlighting the critical role of synthetic parameters in directing phase transformations within the $(Bi_2)_m(Bi_2Te_3)_n$ homologous series. *In situ* heating studies provided detailed structural insights, showing that heating promotes the transformation of BiTe and $Bi_4Te_5$ nanosheets into the more stable $Bi_2Te_3$ at ~340 °C, followed at higher temperatures by oxidation in air or preferential Te sublimation under vacuum. The systematic changes of diffraction patterns and EDS elemental analysis provide direct crystallographic and compositional evidence for these temperature-dependent transformations, confirming the close structural and thermal relationship among these phases. Our findings not only deepen the fundamental understanding of phase stability, sublimation behavior, and structural transformations in the Bi-Te system at the nanoscale, but also offer important insights that may guide the future development of thermoelectric, spintronic, and topological systems. Ultimately, achieving precise control over nanosheet size, morphology, and phase composition represents a meaningful step toward exploring their potential in next-generation electronic and quantum-enabled applications.

## EXPERIMENTAL SECTION

**Chemicals and materials.** Bismuth (Ⅲ) acetate (Bi $(CH_3CO_2)_3$, anhydrous, ⩾99.99%, stored in a nitrogen-filled glovebox), tellurium shots (Te, amorphous, 1-2 mm; 99.999 %; stored in a nitrogen-filled glovebox), diphenyl ditellurid ($(C_6H_5)_2Te_2$, powder, 98%, stored in a nitrogen-filled glovebox) and oleic acid (90%), were purchased from Sigma-Aldrich. Trio-n-octylphosphine (TOP; 97%; stored in a nitrogen-filled glovebox) was from ABCR. Diphenyl ether (⩾99%) and oleylamine (90%) were purchased from Thermo Fisher Scientific. Toluene was purchased from Honeywell. All the chemicals were used as-received without additional purification.

The Trio-n-ctylphosphine-Te (TOP-Te, 1 M) solution was prepared by dissolving Te powder (638 mg, 5.0 mmol) in TOP (5.0 mL) at room temperature until clear. The diphenyl ditellurid-oleylamine (DPD-Te-oleylamine, 0.25 M) solution was prepared by dissolving diphenyl ditelluride (DPD-Te, 512 mg, 1.25 mmol) in oleylamine (5.0 mL) at room temperature in a nitrogen-filled glovebox. The 0.25 M concentration was chosen because higher concentrations resulted in precipitation of DPD-Te at room temperature, while this concentration remained stable and clear. Both solutions were stored in the glovebox until use, and all the syntheses were carried out applying standard air-free Schlenk-line techniques.

**Synthesis of $Bi_2Te_3$ nanosheets.** In a three-necked flask equipped with a condenser, a septum, and a thermocouple in a glass mantle, 96.5 mg of Bi $(CH_3CO_2)_3$ (0.25 mmol), and 2 ml of oleic acid (6.25 mmol) were mixed in 9 ml of diphenyl ether. The mixture was stirred at 110 °C for 10 minutes and subsequently degassed under vacuum at 80 °C for 1 hours. The reaction mixture was then heated to 170 °C under a nitrogen atmosphere. Upon reaching this temperature, 0.2 ml of TOP-Te (0.2 mmol) was swiftly injected, inducing a color change from colorless to black. The resulting nanostructures were purified through precipitation with toluene, followed by centrifugation at 4000 rpm for 3 minutes, repeated three times. The supernatant was discarded, and the purified nanostructures were re-dispersed in toluene for further characterization or storage.

For the “phosphine-free” synthesis approach, 0.8 ml of DPD-Te-oleylamine (0.2 mmol) was used in place of TOP-Te and hot-injected into the solution at 250 °C, with all other reaction conditions remaining the same.

**Structural characterization.** The Transmission Electron Microscopy (TEM) images were performed on a *Talos-L120C* microscope with a thermal emitter operated at an acceleration voltage of 120 kV. The TEM samples were prepared by diluting the nanosheet suspension with toluene, followed by drop casting 10 ml of the suspension on a TEM copper grid coated with a carbon film. The crystal structure of the Bi nanosheets was determined by X-ray Diffraction (XRD) measurements, which were performed on a Panalytical Aeris System with a Bragg–Brentano geometry and a copper anode with an X-ray wavelength of 0.154 nm from the Cu-K$\alpha_1$ line. The samples were prepared by drop-casting 10 ml of the suspended Bi NS solution on a 〈911〉 or 〈711〉 cut silicon substrate. The XRD patterns were fitted using the Pseudo-Voigt peak shape function within the HighScore Plus software. This function, which is a linear combination of Gaussian and Lorentzian components, was chosen because it effectively models both the instrumental broadening (Gaussian contribution) and the sample-induced broadening caused by finite crystallite size and strain (Lorentzian and Gaussian contribution).[41]

Atomic force microscopy (AFM) measurements were performed with AFM from Park Systems XE-100 in non-contact mode.

Scanning transmission electron microscopy (STEM) and electron energy loss spectroscopy (EELS) measurements were performed using a probe-corrected JEOL JEM-ARM200F NeoARM with a cold field emission gun and a Gatan Continuum EELS. All measurements were done at 200 kV acceleration voltage. The full width at half maximum of the zero-loss peak is 0.27 eV with the dispersion set to 0.015 eV/channel. To improve reliability, the first Dual EELS channel was used to track the position of the ZLP while the second channel was shifted to increase the signal-to-noise ratio due to higher acquisition times. During the experimental process, we searched many regions on the TEM grid to identify areas where the nanosheets were well-separated, not stacking or touching each other. We selected these kinds of regions to ensure that our analysis was performed on individual nanosheets.

**Temperature-dependent XRD and ex situ annealing study of phase transformation.** The crystal structures of BiTe and $Bi_4Te_5$ nanosheets subjected to different thermal treatments were characterized by X-ray diffraction (XRD). The measurements were performed in air using a PANalytical Aeris diffractometer equipped with a Bragg–Brentano geometry and a Cu anode. Samples were prepared by drop-casting 10 ml of a dispersion of BiTe or $Bi_4Te_5$ nanosheets onto ⟨911⟩- or ⟨711⟩-oriented silicon substrates and drying under ambient conditions. The substrates were then directly heated in ambient air for 5 min at selected temperatures, followed by natural cooling to room temperature prior to XRD measurements. This heating–cooling–measurement procedure was repeated for different annealing temperatures to investigate the temperature-induced phase evolution of the nanosheets systematically.

The *ex situ* heating experiments were performed on BiTe- and $Bi_4Te_5$-containing nanosheets directly supported on Ni grids under ambient air. The samples were heated in a muffle furnace under ambient conditions at a rate of 10 °C/min to the designated temperature, held for 5 min, and then cooled naturally to room temperature. Subsequently, the nanosheets were characterized using STEM and EDS to investigate the structural and compositional changes.

**The *in situ* annealing and characterization in TEM.** Nanocrystals (NCs) were characterized by TEM and EDS using a probe-corrected JEOL JEM-ARM200F NeoARM transmission electron microscope equipped with a cold field emission gun, a 100 $mm^2$ JEOL EDS detector, and a Gatan

OneView IS camera. The microscope was operated at 200 kV under high-vacuum conditions (1.8 × $10^{-5}$ Pa) using a Protochips Fusion *in situ* heating holder.

For *in situ* annealing, dispersions of BiTe and $Bi_4Te_5$ nanosheets were drop-cast onto a pre-calibrated heating chip coated with an approximately 40 nm thick amorphous silicon nitride membrane. The temperature was ramped from 25 °C to selected target temperatures at a rate of ~1 °C $s^{-1}$. After reaching each target temperature, the sample was cooled at ~10 °C $s^{-1}$ to room temperature for EDS measurements. Subsequently, the sample was reheated to the previously reached target temperature at the higher rate and then slowly heated further to observe structural transitions. This procedure was repeated until the maximum target temperature was reached. TEM imaging was performed continuously throughout the experiment, whereas EDS measurements were acquired exclusively at room temperature.

**Density functional theory (DFT) calculations**. We performed density functional theory calculations using the ORCA code with the B3LYP hybrid functional and the DEF2-TZVP basis set.[29, 30]

ASSOCIATED CONTENT

**Supporting Information**

The Supporting Information is available free of charge at https://pubs.acs.org

TEM images, STEM-EDX analysis, temperature-dependent XRD Analysis of phase evolution and *in situ* annealing and characterization in TEM. (PDF and Videos)

AUTHOR INFORMATION

**Corresponding Author**

fa-gui.he@uni-rostock.de

christian.klinke@uni-rostock.de

**Author Contributions**

The manuscript was written through the contributions of all authors. All authors have given approval to the final version of the manuscript.

**Acknowledgements**

F.G.He acknowledges Dr. Brindhu Malani S, Dr. Eugen Klein and Urvi Parekh for their valuable technical support and helpful discussions. F.G.He. also acknowledges Xuejie Song for technical assistance with the graphical illustrations. Deutsche Forschungsgemeinschaft (DFG, German Research Foundation) is acknowledged for funding of HE8838/3-1, project number 525993990, also acknowledges the European Regional Development Fund of the European Union for funding the X-ray diffractometer (GHS-20-0036/P000379642) and the DFG for funding the transmission electron microscopes Jeol JEM-ARM200F NeoARM (INST 264/161-1 FUGG) and Thermo Fisher Talos L120C (INST 264/188-1 FUGG).

**Conflict of Interest**

The authors declare no conflict of interest.

REFERENCES

(1) Liang, Y.; Wang, W.; Zeng, B.; Zhang, G.; Huang, J.; Li, J.; Li, T.; Song, Y.; Zhang, X. Raman scattering investigation of Bi2Te3 hexagonal nanoplates prepared by a solvothermal process in the absence of NaOH. *J. Alloys Compd.* **2011**, *509*, 5147-5151.

(2) Lu, W.; Ding, Y.; Chen, Y.; Wang, Z. L.; Fang, J. Bismuth Telluride Hexagonal Nanoplatelets and Their Two-Step Epitaxial Growth. *J. Am. Chem. Soc.* **2005**, *127*, 10112-10116.

(3) Scheele, M.; Oeschler, N.; Meier, K.; Kornowski, A.; Klinke, C.; Weller, H. Synthesis and Thermoelectric Characterization of Bi2Te3 Nanoparticles. *Adv. Funct. Mater.* **2009**, *19*, 3476-3483.

(4) Zhang, H.; Liu, C.-X.; Qi, X.-L.; Dai, X.; Fang, Z.; Zhang, S.-C. Topological insulators in Bi2Se3, Bi2Te3 and Sb2Te3 with a single Dirac cone on the surface. *Nat. Phys.* **2009**, *5*, 438-442.

(5) Sokolikova, M. S.; Sherrell, P. C.; Palczynski, P.; Bemmer, V. L.; Mattevi, C. Room-temperature growth of colloidal Bi2Te3 nanosheets. *Chem. Commun.* **2017**, *53*, 8026-8029.

(6) Bae, C.; Böhnert, T.; Gooth, J.; Lim, S.; Lee, S.; Kim, H.; Heimann, S.; Schulz, S.; Shin, H.; Nielsch, K. Thermopower engineering of Bi2Te3 without alloying: the interplay between nanostructuring and defect activation. *Semicond. Sci. Technol.* **2014**, *29*, 064003.

(7) Ritter, J. J. A Novel Synthesis of Polycrystalline Bismuth Telluride. *Inorg. Chem.* **1994**, *33*, 6419-6420.

(8) Ritter, J. J.; Maruthamuthu, P. Synthesis of Polycrystalline Bismuth Telluride by a Metal-Organo Complex Method. *Inorg. Chem.* **1995**, *34*, 4278-4280.

(9) Groshens, T. J.; Gedridge, R. W., Jr.; Lowe-Ma, C. K. Room-Temperature MOCVD of Sb2Te3 Films and Solution Precipitation of M2Te3(M = Sb, Bi) Powders via a Novel N,N-Dimethylaminotrimethylsilane Elimination Reaction. *Chem. Mater.* **1994**, *6*, 727-729.

(10) Xiao, F.; Yoo, B.; Lee, K. H.; Myung, N. V. Synthesis of Bi2Te3 Nanotubes by Galvanic Displacement. *J. Am. Chem. Soc.* **2007**, *129*, 10068-10069.

(11) Jiang, Y.; Zhu, Y.-J. Bi2Te3 nanostructures prepared by microwave heating. *J. Cryst. Growth* **2007**, *306*, 351-355.

(12) Salavati-Niasari, M.; Bazarganipour, M.; Davar, F. Hydrothermal preparation and characterization of based-alloy Bi2Te3 nanostructure with different morphology. *J. Alloys Compd.* **2010**, *489*, 530-534.

(13) Li, G.-R.; Zheng, F.-L.; Tong, Y.-X. Controllable Synthesis of Bi2Te3 Intermetallic Compounds with Hierarchical Nanostructures via Electrochemical Deposition Route. *Cryst. Growth Des.* **2008**, *8*, 1226-1232.

(14) Deng, Y.; Cui, C.-W.; Zhang, N.-L.; Ji, T.-H.; Yang, Q.-L.; Guo, L. Fabrication of bismuth telluride nanotubes via a simple solvothermal process. *Solid State Commun.* **2006**, *138*, 111-113.

(15) Hosokawa, Y.; Tomita, K.; Takashiri, M. Growth of single-crystalline Bi2Te3 hexagonal nanoplates with and without single nanopores during temperature-controlled solvothermal synthesis. *Sci. Rep.* **2019**, *9*, 10790.

(16) Kaibe, H.; Tanaka, Y.; Sakata, M.; Nishida, I. Anisotropic galvanomagnetic and thermoelectric properties of n-type Bi2Te3 single crystal with the composition of a useful thermoelectric cooling material. *J. Phys. Chem. Solids* **1989**, *50*, 945-950.

(17) He, F.; Klein, E.; Bartling, S.; Saeidpour, S.; Corzilius, B.; Lesyuk, R.; Klinke, C. Template-Mediated Formation of Colloidal Two-Dimensional Tin Telluride Nanosheets and the Role of the Ligands. *J. Phys. Chem. C* **2022**, *126*, 20498-20504.

(18) Bos, J. W. G.; Zandbergen, H. W.; Lee, M. H.; Ong, N. P.; Cava, R. J. Structures and thermoelectric properties of the infinitely adaptive series (Bi2)m(Bi2Te3)n. *Phys. Rev. B* **2007**, *75*, 195203.

(19) Bos, J. W. G.; Faucheux, F.; Downie, R. A.; Marcinkova, A. Phase stability, structures and properties of the (Bi2)m (Bi2Te3)n natural superlattices. *J. Solid State Chem.* **2012**, *193*, 13-18.

(20) Ouyang, Y.; Zhang, M.; Li, C.; Li, X.; Xie, S.; Yan, F.; Ge, H.; Li, Z.; Tong, Q.; Poudeu, P. F. P.; et al. Interactions between Functional Units Inducing the Evolution of Electronic Band

Structure and Improved Thermoelectric Performance of n-Type (Bi2)x(Bi2Te3)y Pseudosuperlattices. *ACS Appl. Energy Mater.* **2024**, *7*, 12048-12058.

(21) Cruz-Gandarilla, F.; Vigil-Galán, O.; Cabañas-Moreno, J. G.; Sastré-Hernández, J.; Roy, F. Structural and microstructural characterization of Bi2Te3 films deposited by the close space vapor transport method using scanning electron microscopy and X-ray diffraction techniques. *Thin Solid Films* **2012**, *520*, 3865-3870.

(22) Concepción, O.; Galván-Arellano, M.; Torres-Costa, V.; Climent-Font, A.; Bahena, D.; Manso Silván, M.; Escobosa, A.; de Melo, O. Controlling the Epitaxial Growth of Bi2Te3, BiTe, and Bi4Te3 Pure Phases by Physical Vapor Transport. *Inorg. Chem.* **2018**, *57*, 10090-10099.

(23) Pathak, R.; Joseph, A.; Dutta, P.; Joseph, B.; Duhan, J.; Chandra, S.; Narayana, C.; Pal, K.; Biswas, K. Impact of Pressure on Metavalent Bonding in BiTe Influencing Electronic Topological Transitions. *Angew. Chem. Int. Ed.* **2025**, *64*, e202422652.

(24) He, F.; Grisan Qiu, Y. Y.; Mearini, S.; Feyer, V.; Oldenburg, K.; Lesyuk, R.; Klinke, C. Tunable Colloidal Synthesis Enabling μ-ARPES on Individual Bismuth Nanocrystals. *Small* **2026**, *22*, e12032. .

(25) Abellan, P.; Parent, L. R.; Al Hasan, N.; Park, C.; Arslan, I.; Karim, A. M.; Evans, J. E.; Browning, N. D. Gaining Control over Radiolytic Synthesis of Uniform Sub-3-nanometer Palladium Nanoparticles: Use of Aromatic Liquids in the Electron Microscope. *Langmuir* **2016**, *32*, 1468-1477.

(26) Karim, A. M.; Al Hasan, N.; Ivanov, S.; Siefert, S.; Kelly, R. T.; Hallfors, N. G.; Benavidez, A.; Kovarik, L.; Jenkins, A.; Winans, R. E.; et al. Synthesis of 1 nm Pd Nanoparticles in a Microfluidic Reactor: Insights from in Situ X-ray Absorption Fine Structure Spectroscopy and Small-Angle X-ray Scattering. *J. Phys. Chem. C* **2015**, *119*, 13257-13267.

(27) Ortiz, N.; Skrabalak, S. E. Manipulating Local Ligand Environments for the Controlled Nucleation of Metal Nanoparticles and their Assembly into Nanodendrites. *Angew. Chem. Int. Ed.* **2012**, *51*, 11757-11761.

(28) Mishra, M.; Patil, N. N.; Zubair, M.; Kapuria, N.; Lebedev, V.; Adegoke, T. E.; Ryan, K. M.; Singh, S. Phosphine free synthesis of copper telluride nanocrystals in 1D and 2D shapes using Dipehylditelluride (DPDTe) as an air-stable source. *Nanotechnology* **2022**, *33*, 305602.

(29) Neese, F. Software Update: The ORCA Program System—Version 6.0. *WIREs Computational Molecular Science* **2025**, *15*, e70019.

(30) Neese, F. The ORCA program system. *WIREs Computational Molecular Science* **2012**, *2*, 73-78.

(31) La, H.; Brokkelkamp, A.; van der Lippe, S.; ter Hoeve, J.; Rojo, J.; Conesa-Boj, S. Edge-induced excitations in Bi2Te3 from spatially-resolved electron energy-gain spectroscopy. *Ultramicroscopy* **2023**, *254*, 113841.

(32) Nascimento, V. B.; de Carvalho, V. E.; Paniago, R.; Soares, E. A.; Ladeira, L. O.; Pfannes, H. D. XPS and EELS study of the bismuth selenide. *J. Electron. Spectrosc. Relat. Phenom.* **1999**, *104*, 99-107.

(33) Cha, J. J.; Koski, K. J.; Huang, K. C. Y.; Wang, K. X.; Luo, W.; Kong, D.; Yu, Z.; Fan, S.; Brongersma, M. L.; Cui, Y. Two-Dimensional Chalcogenide Nanoplates as Tunable Metamaterials via Chemical Intercalation. *Nano Lett.* **2013**, *13*, 5913-5918.

(34) Gierlotka, W. A new thermodynamic description of the binary Bi-Te system using the associate solution and the Wagner-Schottky models. *Calphad* **2018**, *63*, 6-11.

(35) Drzewowska, S.; Onderka, B. Different Approach to Thermodynamic Description of Bi-Te Binary System. *J. Phase Equilib. Diffus.* **2023**, *44*, 429-444.

(36) Buha, J.; Gaspari, R.; Del Rio Castillo, A. E.; Bonaccorso, F.; Manna, L. Thermal Stability and Anisotropic Sublimation of Two-Dimensional Colloidal Bi2Te3 and Bi2Se3 Nanocrystals. *Nano Lett.* **2016**, *16*, 4217-4223.

(37) Huang, H.; Luan, W.-l.; Tu, S.-t. Influence of annealing on thermoelectric properties of bismuth telluride films grown via radio frequency magnetron sputtering. *Thin Solid Films* **2009**, *517*, 3731-3734.

(38) Samanta, M.; Pal, K.; Waghmare, U. V.; Biswas, K. Intrinsically Low Thermal Conductivity and High Carrier Mobility in Dual Topological Quantum Material, n-Type BiTe. *Angew. Chem. Int. Ed.* **2020**, *59*, 4822-4829.

(39) Subrati, A.; Kim, Y.; Al Wahedi, Y.; Tzitzios, V.; Alhassan, S.; Kim, H. J.; Lee, S.; Sakellis, E.; Boukos, N.; Stephen, S.; et al. Monitoring the multiphasic evolution of bismuth telluride nanoplatelets. *CrystEngComm* **2020**, *22*, 7918-7928.

(40) Eschbach, M.; Lanius, M.; Niu, C.; Młyńczak, E.; Gospodarič, P.; Kellner, J.; Schüffelgen, P.; Gehlmann, M.; Döring, S.; Neumann, E.; et al. Bi1Te1 is a dual topological insulator. *Nat. Commun.* **2017**, *8*, 14976.

(41) De Keijser, T. H.; Langford, J. I.; Mittemeijer, E. J.; Vogels, A. B. P. Use of the Voigt function in a single-line method for the analysis of X-ray diffraction line broadening. *J. Appl. Crystallogr.* **1982**, *15*, 308-314.